\documentstyle[emulateapj]{article}
\input epsf.sty

\slugcomment{Accepted for publication in ApJL}

\newcommand\F{{\cal F}}
\newcommand\al{\alpha}
\renewcommand\th{\theta}

\newcommand\Hunits{\mbox{km s$^{-1}$ Mpc$^{-1}$}}

\newcommand\hMpc{\mbox{$h^{-1}\,$Mpc}}
\newcommand\hMsun{\mbox{$h^{-1}\,M_\odot$}}
\newcommand\avg[1]{\langle{#1}\rangle}
\newcommand\src{{\rm src}}
\newcommand\lens{{\rm lens}}
\newcommand\med{{\rm med}}
\newcommand\rr{r_{-2}}
\newcommand\CC{C_{-2}}
\newcommand\fcore{f_{\rm core}}

\newcommand\refeq[1]{eq.~(\ref{eq:#1})}

\makeatletter
\newenvironment{tablehere}
  {\def\@captype{table}}
  {}
\newenvironment{figurehere}
  {\def\@captype{figure}}
  {}
\makeatother

\begin{document}
\title{Lensing Constraints on the Cores of Massive Dark Matter Halos}
\author{Charles R.\ Keeton\altaffilmark{1} and Piero Madau\altaffilmark{2}}
\altaffiltext{1}{Steward Observatory, University of Arizona,
  933 N.\ Cherry Ave., Tucson AZ 85721; email ckeeton@as.arizona.edu}
{\altaffiltext{2}{Department of Astronomy and Astrophysics,
  University of California, Santa Cruz, CA 95064; email pmadau@ucolick.org}

\begin{abstract}
The statistics of wide-separation ($6\arcsec < \th < 15\arcsec$)
gravitational lenses constrain the amount of mass in the cores of
dark matter halos on group and cluster mass scales.  For a family
of halo models with a central cusp $\rho \propto r^{-\al}$ ($1.0
\le \al \le 1.9$), the lack of wide-separation lenses in the
large Cosmic Lens All-Sky Survey yields an upper limit on the
fraction of the halo mass that is contained within $\sim$4\% of
the virial radius, $\fcore<0.023$ (95\% confidence level, LCDM).
This limit offers an important test of the cold dark matter
paradigm. While the halo profiles derived from numerical
simulations appear to be consistent with this upper limit, larger
surveys currently underway such as the 2dF and SDSS should detect
wide-separation lenses and thus provide a measurement of the core
mass fraction in massive dark matter halos.
\end{abstract}

\section{Introduction}

Cuspy dark matter halos are a robust prediction of the cold dark
matter (CDM) paradigm. Numerical simulations have shown that there
is a ``universal'' density profile for dark matter halos with the
form $\rho \propto r^{-1}\,(r_s+r)^{-2}$, where $r_s$ is a scale
length (Navarro, Frenk \& White 1997, hereafter NFW).  This
result does not depend on particular cosmogonies or on specific
initial conditions or formation histories (Huss, Jain \&
Steinmetz 1999a,b). There is some disagreement about the very
inner regions of halos; recent simulations have suggested that
the central cusp may be steeper than $r^{-1}$ (e.g., Moore et
al.\ 1998, 1999) and may depend on the halo mass (Jing \& Suto
2000).

It is important to compare the predicted halo profiles with
observational data to test the CDM paradigm.  Galaxy rotation
curves can be used to probe dark matter halos (e.g., Flores \&
Primack 1994; McGaugh \& de Blok 1998), but the implications are
not clear because beam smearing can make HI rotation curves
appear shallower than they really are (van den Bosch \& Swaters
2000). Gravitational lensing offers an additional test of halo
profiles. More than 60 multiply-imaged quasars and radio sources
are known (see Falco et al.\ 1999), and with image separations
of a few arcseconds they primarily probe halos on galaxy mass
scales. These lenses can be used to constrain CDM halos (e.g.,
Rusin \& Ma 2000), but the test is complicated by the necessity
of including baryons and their effects on dark matter halos.  At
least 24 giant lensed arcs are also known (see Williams, Navarro
\& Bartelmann 1999), and with image separations of several tens
of arcseconds or larger they probe rich clusters.

Lenses with intermediate image separations, $\th \sim 10\arcsec$,
probe halos at intermediate masses.  There is just one confirmed
lens (Q~0957+561, Walsh, Carswell \& Weymann 1979) and one good
candidate (RX~J0921+4529, Mu\~noz et al.\ 2000) with $\th > 6
\arcsec$, and each is produced by a galaxy in a cluster.  No
systematic lens survey has found any lenses in this regime, and
this result has been used to place limits on cosmological
parameters (e.g., Cen et al.\ 1994; Kochanek 1995; Wambsganss et
al.\ 1995; Maoz et al.\ 1997; Mortlock \& Webster 2000). Instead,
in this {\it Letter\/} we propose to use the statistics of
wide-separation lenses as a new test of the structure of dark
matter halos (see Flores \& Primack 1996).  We consider a range
of halo profiles suggested by recent CDM simulations, and compare
our model predictions with the latest and largest lens survey,
the Cosmic Lens All-Sky Survey.

\section{Cuspy halos}

To obtain a family of halos with a range of central cusps, we
adopt a generalization of the NFW profile (see Jing \& Suto 2000;
Wyithe, Turner \& Spergel 2000),
\begin{equation} \label{eq:cuspy}
  \rho(r) = {\rho_s \over (r/r_s)^\al (1+r/r_s)^{3-\al}}\ ,
\end{equation}
where $r_s$ is a scale length and $\rho_s$ is a characteristic
density.  The density scales as $\rho \propto r^{-\al}$ for small
$r$ and $\rho \propto r^{-3}$ for large $r$, so it reduces to the
NFW profile when $\al=1$.  The characteristic density $\rho_s$ is
given by
\begin{equation}
  \rho_s =\rho_{\rm crit}(z)\, {200(3-\al)\,(r_{200}/r_s)^{\al-1} \over
    {}3\, _2 F_1 [ 3-\al, 3-\al, 4-\al, -r_{200}/r_s]}\ ,
\end{equation}
where ${} _2 F_1(a,b,c,z)$ is a hypergeometric function, and
$r_{200}$ is the radius within which the mean density is 200
times the critical density of the universe $\rho_{\rm crit}$.
Moore et al.\ (1998, 1999) advocate a profile of the form $\rho
\propto x^{-1.5} (1+x^{1.5})^{-1}$ where $x=r/r_s$. Although this
differs slightly from a cuspy profile with $\al=1.5$, the two
profiles are similar at small radii and differ by $\lesssim$15\%
in the enclosed inner mass, which we will argue below is the most
important quantity in determining the number of lenses.  We do
not consider the Moore profile further because there is no simple
family of models connecting it to the NFW model. It is often
convenient to replace the scale radius of a cuspy halo with a
``concentration'' parameter. For NFW ($\al=1$) halos, Navarro et
al.\ (1997) define $C_{\rm NFW} \equiv r_{200}/r_s$.  For the
generalization to $\al \ne 1$, we define $\CC \equiv r_{200}/\rr$
in terms of the radius $\rr$ at which the logarithmic slope of
the density profile is $-2$, which is equivalent to $C_{\rm NFW}$
for $\al=1$.\footnote{For $\al \ne 1$, $\CC$ differs from the
parameter $C_s \equiv r_{200}/r_s$ used by Wyithe et al.\ (2000).
We believe that $\CC$ is the better generalization to $\al \ne 1$
because the radius $\rr$ has physical significance, while $r_s$
does not; in any case, the alternate parameters have a simple
relation in cuspy halos, $C_s = (2-\al) \CC$.}

A cuspy halo is fully specified by its mass, redshift, cusp
slope, and scale radius or concentration, but these parameters
are not all independent.  For NFW halos, the concentration is
correlated with the mass and redshift in a way that appears to
contain information about the formation epoch of the halo (e.g.,
Navarro et al.\ 1997).  However, there is a scatter in
concentrations at fixed mass and redshift that may reflect
differences in formation histories (e.g., Bullock et al.\ 1999;
Klypin et al.\ 2000).  The scatter is important for our
calculations because lensing is very sensitive to concentration;
more concentrated halos are much more efficient lenses.  We use
results from the simulations by Bullock et al.\ (1999), namely a
scatter that is consistent with a log-normal distribution with
standard deviation $\Delta(\log\CC) \simeq 0.18$, and a median
concentration that scales with mass and redshift as $\med(\CC)
\propto M^{-1/9} (1+z)^{-1}$ (Bullock et al. 1999; Navarro \&
Steinmetz 2000).  These trends were derived from fits of NFW
($\al=1$) profiles to simulated halos, but because their origin
is physical we assume that they apply to other values of $\al$ as
well.

\section{Lens theory}

To compute the total number of lenses expected to be produced by
a population of halos it is sufficient to use spherical halos.
Departures from spherical symmetry mainly affect the relative
numbers of two-image and four-image lenses, which we do not
differentiate. Wyithe et al.\ (2000) and Li \& Ostriker (2000)
discuss the lensing properties of cuspy halos of the form given
in \refeq{cuspy}, but their formalism can be simplified so the
deflection is written as
\begin{equation} \label{eq:def}
  \phi_{,R}(R) = {R \over \pi \Sigma_{\rm cr}} \int_{R}^{\infty}
    {M(r) \over r^2 \sqrt{r^2-R^2}}\,dr\,,
\end{equation}
where $M(r)$ is the mass inside radius $r$ and $\Sigma_{\rm cr} =
(c^2 D_s) / (4 \pi G D_l D_{ls})$ is the critical surface density
for lensing (e.g., Schneider, Ehlers \& Falco 1992).  Here $D_l$
and $D_s$ are angular diameter distances to the lens and source,
respectively, and $D_{ls}$ is the angular diameter distance from
the lens to the source. The positions of lensed images are found
by solving the lens equation $u = R - \phi_{,R}(R)$, where $u$ is
the impact parameter of the source relative to the lens.  The
magnification of an image at $R$ is then $\mu =
[1-\phi_{,R}/R]^{-1} [1-\phi_{,RR}]^{-1}$, where $\phi_{,RR} =
d(\phi_{,R})/dR$.

The number of lenses expected to be found in a survey depends on
the optical depth for lensing as well as on the ``magnification
bias,'' which accounts for lenses that are intrinsically fainter
than the flux limit of the survey but are brought into the sample
by the magnification (e.g., Turner, Ostriker \& Gott 1984).  The
number of lenses with a total flux greater than $S$ expected to
be found in a survey with selection functions described by $\F$
is then
\begin{eqnarray}
  N_\lens(>\!S;\F) &=& {1 \over 4\pi} \int dz_s \int dV
    \int dM\,{dn \over dM} \\
  && \times \int_{\rm mult} du\,2\pi u\,
    \F(u)\,{dN_\src(>\!S/\mu) \over dz_s}\ , \nonumber
\end{eqnarray}
where $z_s$ is the source redshift, $dV$ is the comoving volume
element, and $dn/dM$ is the mass function of halos.  Also, $u$ is
the angular position of the source relative to the lens center,
and $\mu$ is the total magnification of that source; the $u$
integral extends over the range of impact parameters that produce
multiple images. Finally, $[dN_\src(>\!S)/dz_s]\,dz_s$ is the
number of sources brighter than flux $S$ that lie in the redshift
range $z_s$ to $z_s+dz_s$. The factor ${\cal F}(u)$ indicates
whether a lens associated with a source at $u$ would be detected
given the selection functions.

We follow Narayan \& White (1988), Kochanek (1995), and Porciani
\& Madau (2000) and compute lens statistics using a mass function
of dark matter halos given by Press-Schechter theory combined
with the spherical collapse model. Numerical simulations suggest
that this mass function overestimates the number of halos below
$\sim\!10^{14}\,\hMsun$ and underestimates the number of halos
above this mass (e.g., Jenkins et al.\ 2000). Because we are
interested in the high-mass end of the mass function, our results
should slightly underestimate the number of lenses, and lead to
conclusions that are conservative. We compute the CDM power
spectrum using the fitting formula given by Eisenstein \& Hu
(1999).  We present results for two different flat cosmologies:
SCDM, with matter density $\Omega_M = 1$ and Hubble constant $H_0
= 50\,\Hunits$; and LCDM, with $\Omega_M = 0.3$, $\Omega_\Lambda
= 0.7$, and $H_0 = 65\,\Hunits$.  From the abundance of clusters
we set the variance of mass fluctuations on $8\,\hMpc$ scales to
be $\sigma_8 = 0.52$ for SCDM and $\sigma_8 = 0.93$ for LCDM
(Eke, Cole \& Frenk 1996).

Li \& Ostriker (2000) have recently performed a similar
calculation. However, they use a different power spectrum, omit
magnification bias, and do not consider a scatter in
concentrations, so their quantitative results are somewhat
different from ours.

\section{The CLASS lens survey}

To test the models we must compare predicted lens statistics with
a well-defined observational sample. The Cosmic Lens All-Sky
Survey (CLASS; e.g., Browne \& Myers 2000) is the largest statistically
homogeneous search for gravitational lenses. The sample comprises
$10,499$ flat-spectrum radio sources whose flux distribution can
be described as a power law $dN_\src/dS \propto S^{-2.1}$ (see
Rusin \& Tegmark 2000).  The survey includes 18 gravitational
lenses with image separations $\th < 3\arcsec$.  An explicit
search for lenses with image separations $6\arcsec < \th <
15\arcsec$ has found no lenses (Phillips et al.\ 2000).

The redshift distribution of the full CLASS sample is not known.
Marlow et al.\ (2000) report redshifts for a small subsample of 27
sources.  They find a mean redshift of $\avg{z}=1.27$, which is
comparable to that found in other radio surveys at comparable
fluxes. We present results assuming that the full sample has the
same source redshift distribution as the spectroscopic subsample
of Marlow et al.\ (2000), but our results are not substantially
different if we simply place all sources at the mean redshift of
the spectroscopic subsample, so the source redshift distribution
is not a significant source of uncertainty.

\section{Results}

We focus on wide-separation lenses, $6\arcsec < \th < 15\arcsec$,
which are not present in the CLASS sample and correspond to halos
on the mass scale of groups and clusters.  The exact bounds on
the image separation have no intrinsic significance but are
chosen to correspond to the analysis of the CLASS survey data
(see Phillips et al.\ 2000). We compute the number of lenses for
particular values of the cusp slope and median concentration; we
consider a wide range of models with $1.0 \le \al \le 1.9$ and $2
\le \med(\CC) \le 11$. Models with $\med(\CC) > 11$ are ruled out
at 99\% confidence or better for all values of $\al$ that we
consider.

\begin{figurehere}
\centerline{\epsfxsize=3.0in \epsfbox{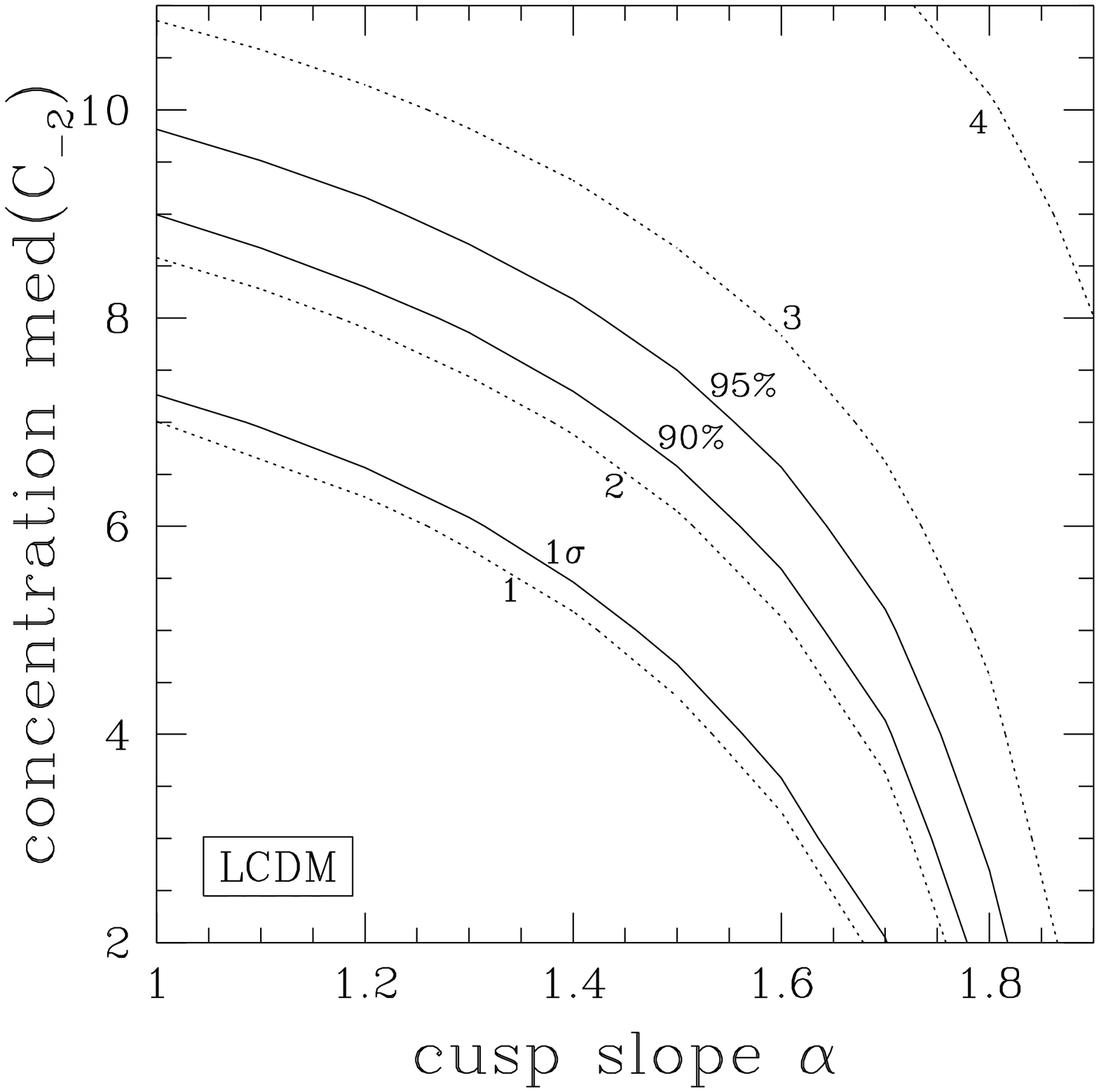}}
\caption{\footnotesize
The number of wide-separation lenses expected in the CLASS survey
as a function of the cusp slope and concentration parameter, for
the LCDM cosmology.  The dotted curves show contours drawn at 1,
2, 3, and 4 lenses, and the solid curves show the $1\sigma$, 90\%,
and 95\% confidence upper limits.
}
\end{figurehere}
\vspace{0.2cm}

Figure 1 shows the expected number of wide-separation lenses as a
function of $\al$ and $\CC$.  The results are not easy to
interpret for two reasons.  First, there is some ambiguity in the
meaning of the concentration parameter when halos are allowed to
have arbitrary cusps (see \S 5 of Wyithe et al.\ 2000 for a
detailed discussion).  Second, lensing is not very sensitive to
the cusp slope and the concentration separately, but depends
mainly on the amount of mass in the central regions of halos.  It
is more instructive then to plot the number of lenses versus the
``core mass fraction'' $\fcore$, or the fraction of the total
mass of the halo that is contained within some small fiducial
radius, as in Figure 2.  The best fiducial radius is the one that
minimizes the scatter in the relation between the core mass
fraction and the number of lenses.\footnote{Specifically, we
minimize the scatter in the log-log relation, including all
models with $1.0 \le \al \le 1.7$ and $2 \le \med(\CC) \le 11$
that predict at least 0.1 lenses.} It is $0.045\,r_{200}$ for
SCDM and $0.032\,r_{200}$ for LCDM, which is approximately equal
to the mean Einstein radius of lenses with $6\arcsec < \th <
15\arcsec$.  Because of systematic trends and the scatter in
concentrations, the core mass fraction is measured for a median
halo of mass $10^{14}\,\hMsun$ at redshift $z=0$, which is the
mass scale relevant for wide-separation lenses.

\begin{figurehere}
\centerline{\epsfxsize=3.0in \epsfbox{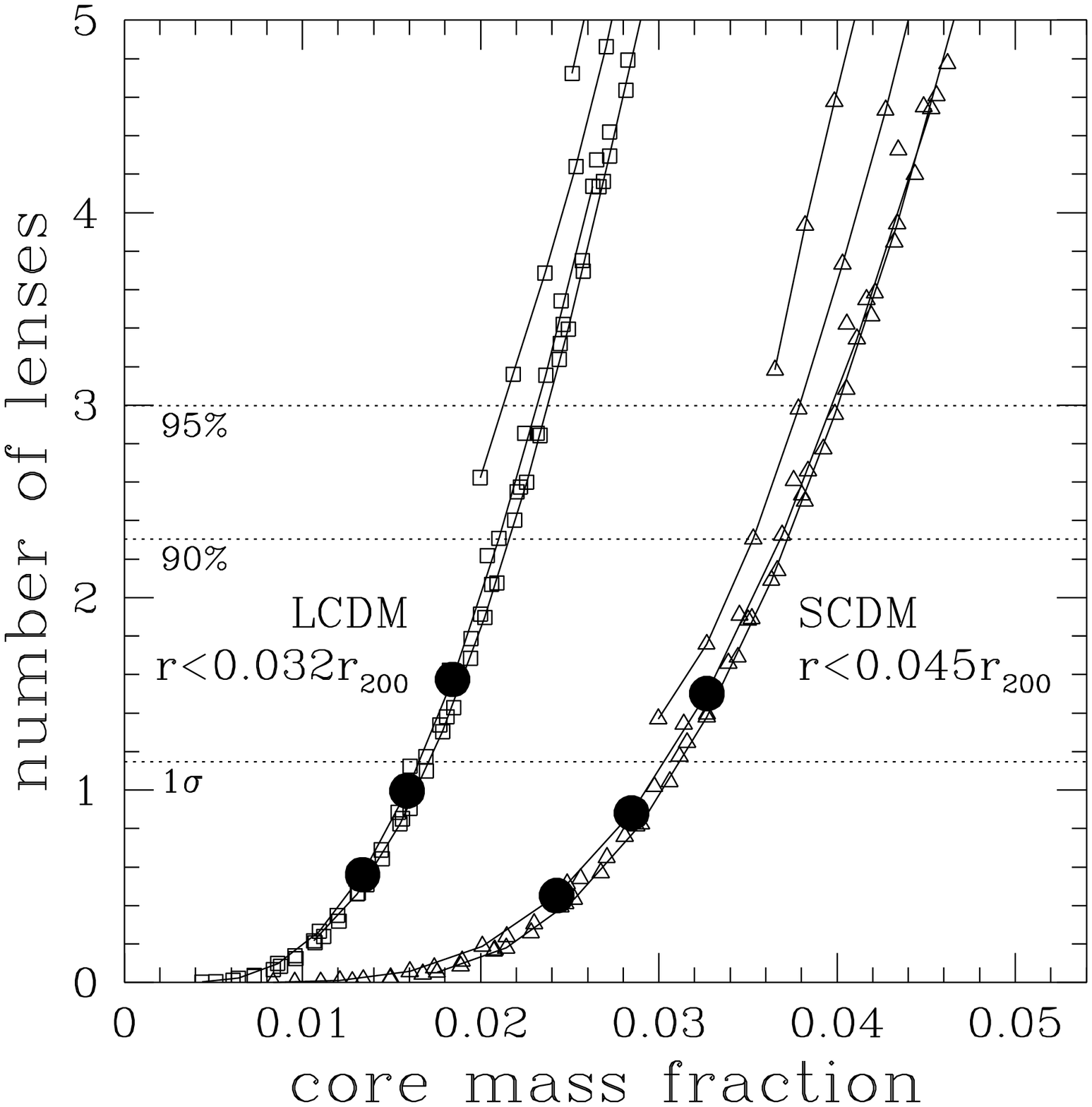}}
\caption{\footnotesize
The number of wide-separation lenses plotted versus the fraction
of halo mass that is contained within a small fiducial radius
(indicated in the label).  Each point indicates a model with
specific values of $\al$ and $\med(\CC)$, with squares (triangles)
for LCDM (SCDM).  The filled circles indicate NFW ($\al=1$) models
with $\med(\CC)=6$, 7, and 8.  Lines connect models with $\al=1$
and 1.5 (in the main relations), and $\al=1.8$ and 1.9 (offset
above the main relations).  The horizontal lines indicate the
$1\sigma$, 90\%, and 95\% confidence upper limits on the number
of lenses.
}
\end{figurehere}
\vspace{0.2cm}

The number of predicted lenses is strongly correlated with the
core mass fraction, and this relation explains most of the
dependence of the number of lenses on the cusp slope and the
concentration.  This result is surprising in its simplicity,
because lensing actually depends on the projected mass
distribution and because it is not obvious that lensing should
measure the {\it fraction\/} of mass that is contained within a
fiducial radius expressed as a {\it fraction\/} of the halo's
radius.  Indeed, for these reasons the relation between lensing
and the core mass fraction is not perfect; models with steep
cusps ($\al = 1.8$ or 1.9) fall somewhat off the main relation,
and softened isothermal models (not shown) follow a similar but
slightly different relation.  Nevertheless, the main relation
between the number of lenses and the core mass fraction is quite
tight for a range of cusp slopes and concentrations that is wider
than the range indicated by numerical simulations. Our first
result, then, is that the statistics of wide-separation lenses
offer a surprisingly good method for constraining the fraction of
mass in group and cluster halos contained within a fiducial
radius that is a few percent of $r_{200}$.

The predicted number of lenses admits the following quantitative
interpretation.  The probability of finding $k$ lenses when $N$
are expected is well approximated by the Poisson distribution as
$P(k|N) = N^k e^{-N}/k!$.  Given that CLASS contains no
wide-separation lenses ($k=0$), the upper limits on the predicted
number of lenses are $N < (1.15, 2.30, 3.00)$ at the $(1\sigma,
90\%, 95\%)$ confidence levels. Figures 1 and 2 then yield upper
limits on the core mass fraction, cusp slope, and concentration,
which are summarized in Table 1. For NFW ($\al=1$) halos, the
median concentration of a $10^{14}\,\hMsun$ halo must be less
than about 7 at $1\sigma$ and less than 9 at 90\%. The
simulations by Navarro et al.\ (1997) and Bullock et al.\ (1999)
yield concentrations of slightly less than 6, so they are
consistent with the lens data (see Figure 2).  Halos that are
significantly more concentrated would disagree with the lens
data.  We could compute the limits on the concentration parameter
for other cusp slopes (and Table 1 includes the limits for
Moore-like $\al=1.5$ halos), but we believe it is more
instructive to focus on the core mass fraction.

Finally, we note that modeling all halos as singular isothermal
spheres yields a prediction of 9 (11) wide-separation lenses for
SCDM (LCDM).  Thus, the statistics of wide-separation lenses
strongly exclude the hypothesis that the halos of groups and
clusters are singular isothermal spheres.

\newcommand\tableskip{}
\begin{tablehere}\footnotesize
\begin{center}
{\sc Upper Limits From Wide-Separation Lenses}\\
\begin{tabular}{ccccc}
\tableskip\hline\hline\tableskip
Cosmology & Quantity & $1\sigma$ & 90\% & 95\% \\
\tableskip\hline\tableskip
SCDM & $\fcore$   & 0.031 & 0.037 &  0.040 \\
     & $\CC(1)$   & 7.6   & 9.0   &  9.7   \\
     & $\CC(1.5)$ & 5.6   & 7.3   &  8.1   \\
LCDM & $\fcore$   & 0.017 & 0.021 &  0.023 \\
     & $\CC(1)$   & 7.3   & 9.0   &  9.8   \\
     & $\CC(1.5)$ & 4.7   & 6.6   &  7.5   \\
\tableskip\hline
\end{tabular}
\end{center}
NOTE.---%
$\fcore$ is the fraction of a median halo's mass that is
contained within $0.045\,r_{200}$ for SCDM, or $0.032\,r_{200}$
for LCDM.  $\CC(\al)$ is the median concentration of halos with
cusp slope $\al$.
\end{tablehere}

\section{Conclusions}

We have studied the statistics of wide-separation ($6\arcsec <
\th < 15\arcsec$) lenses produced by dark matter halos with
general cusps of the form $\rho \propto r^{-\al}$ for $1.0 \le
\al \le 1.9$.  In these models the number of expected lenses is
determined almost entirely by the fraction of the halo mass that
is contained within a fiducial radius that is $\sim$4\% of the
virial radius.  Combining our results with the lack of observed
wide-separation lenses in the CLASS lens survey yields an upper
limit on how concentrated halos can be.  The concentration
parameter for the standard NFW profile must be $C \lesssim 7$
($1\sigma$); see Table 1 for more details. Massive halos found in
numerical simulations of CDM cosmogonies have $C \simeq 6$ (e.g.,
Navarro et al. 1997; Bullock et al. 1999) and thus are consistent
with the lensing constraints, but halos cannot be much more
concentrated than that.  Halos with cusps steeper than $r^{-1}$
are also allowed provided that they have somewhat lower
concentrations.  Although we quote limits on the concentration,
we believe that it is more attractive to state our results in
terms of the core mass fraction, e.g., $\fcore<0.023$ at 95\%
confidence for LCDM.  This quantity should be a more robust
prediction of numerical simulations than the slope of the central
density and should allow a simple comparison of our models with
simulations.

Two points are important for interpreting out results.  First,
the scatter in halo properties requires that we quote median
properties.  Any given halo may be more concentrated than the
median and hence violate the mass limits that we have derived for
median halos. In other words, the lensing limits must be applied
statistically, not to individual halos. Second, we have
considered only dark matter models, neglecting any baryons that
might occupy the cores of massive dark matter halos as cD
galaxies in clusters or central ellipticals in groups. This
approach should lead us to conservative conclusions, because any
central baryonic component would only increase the number of
expected lenses and thus aggravate the discrepancy between the
data and the CDM models.

While our results indicate that the lack of wide-separation
lenses in the CLASS survey is perhaps not surprising, they also
suggest that we should expect to discover them soon. In our
models, NFW halos with a median concentration $C=6$ predict 0.6
lenses with $6\arcsec < \th < 15\arcsec$ out of the 10,499 CLASS
targets.  Several new surveys are large enough that we should
expect a few group and cluster lenses; they include 2dF and SDSS,
which should measure the redshifts of more than 25,000 and $10^5$
quasars, respectively. If the new surveys continue to lack
wide-separation lenses, the lensing constraints on the cores of
dark matter halos will become strong enough to question the CDM
paradigm. If, on the other hand, the new surveys do detect
wide-separation lenses, the lenses will allow empirical
measurements of the properties of the cores of massive dark
matter halos.

\acknowledgements
We thank M. Steinmetz and V. Eke for helpful discussions about
dark matter halos, and P. Phillips and I. Browne for information
about the CLASS survey.
Support for this work was provided by NASA through ATP grant
NAG5-4236 and grant AR--06337.10-94A from the Space Telescope
Science Institute (P.M.).


\end{document}